# Excitation of ship waves by a submerged object: new solution to the classical problem

A. V. Arzhannikov* and I. A. Kotelnikov†
*Budker institute of Nuclear Physics SB RAS and*
*Novosibirsk State University*

We have proposed a new method for solving the problem of ship waves excited on the surface of a non-viscous liquid by a submerged object that moves at a variable speed. As a first application of this method, we have obtained a new solution to the classic problem of ship waves generated by a submerged ball that moves rectilinearly with constant velocity parallel to the equilibrium surface of the liquid. For this example, we have derived asymptotic expressions describing the vertical displacement of the liquid surface in the limit of small and large values of the Froude number. The exact solution is presented in the form of two terms, each of which is reduced to one-dimensional integrals. One term describes the "Bernoulli hump" and another term the "Kelvin wedge." As a second example, we considered vertical oscillation of the submerged ball. In this case, the solution leads to the calculation of one-dimensional integral and describes surface waves propagating from the epicenter above the ball.

## I. INTRODUCTION

The problem of ship waves excited by a moving vessel has a long history, beginning with the work of William Thomson (Lord Kelvin) in 1891 [1], which was followed by a detailed study of the waves on deep water [2, 3]. Applying the method of stationary phase discovered by him [4], Kelvin showed that the gravity waves, excited on the surface of the water by a small ship, which moves along a straight line at a constant speed, extend behind the vessel within the wedge angle $\theta_K = \arcsin(1/3) \approx 19.47°$. An elementary explanation of this fact can be found in section §3.10 of monograph [5] by James Lighthill. A key contribution to the development of the theory was made by Thomas Havelock [6–9], Horace Lamb [10, 11], Einar Hogner [12, 13], A. Peters [14], and F. Ursell [15]. In those early theories, the effect a vessel on the water surface was simulated by an external source of pressure, moving with constant velocity and, in the simplest case, symmetrical about the center of the vessel. The main task of the theory at that time was the the calculation of the wave resistance force exerted on the moving vessel. This resistance occurs even in an ideal inviscid liquid due to the excitation of gravity waves on the surface of the liquid; these waves take away part of the mechanical energy of translational motion of the vessel.

Later, probably in connection with the development of submarine Navy, some authors became interested in the problem of excitation of ship waves by an object moving below the surface of the liquid, causing intensive development of numerical methods of calculation. We deliberately do not discuss such methods; the reader can get an idea of them from the publications in the press [16–24] and few declassified technical reports [25–32].

In recent years, we observe an increasing number of publications on the subject of ship waves. Most active are the Chinese researchers, in collaboration with respected European experts [22, 33–41]. Among other things, they consider the impact of the wind and the finite depths of the sea on the ship waves [35]. Analyzing satellite images from Google Earths [42], Marc Rabaud and Frederick Moses recently discovered [43] that the angle of the Kelvin wedge $\theta_K$ decreases approximately in inverse proportion to the speed of the vessel $V$, if the Froude number $F$ is greater than 0.5. They define the Froude number $F = V/\sqrt{gL}$ as the ratio of $V$ to the square root of the product of acceleration of gravity $g$ on vessel length $L$. These authors explain the effect of reducing $\theta_K$ by taking into account the fact that the waves with a wavelength of the order of the ship length should dominate in the spectrum of the the waves excited by a vessel [44]. Somewhat earlier, a similar effect has been seen in the numerical calculations [35]. Discovery by Rabaud and Moisy generated a flurry of new publications [41, 45–53].

Analytical theory of ship wave generation by high-speed underwater objects is developed in less details than that by surface vessels. More than 100 years ago, Lamb [10] and Havelock [7] found a solution for a cylinder that moves parallel to the surface of an inviscid liquid perpendicular to the cylinder axis. Havelock solved three-dimensional problem for a submerged ball [7, 8] and calculated the force of the wave resistance. In final form of his theory [8], he has derived an expression for elevation $\zeta$ (i.e., vertical displacement) of the liquid free surface caused by the motion of the ball with constant speed. His expression has two terms, one of which can be evaluated in analytic form, and the second one includes double integrals. These terms have no clear physical meaning. In particular, they do not vanish if the ball speed $V$ tends to zero, while the elevation of the liquid surface in this case should be zero everywhere. In other words, $\zeta$ is calculated as a result of almost complete mutual reduction of two large terms. For this reason, Havelock formula is hardly suitable for practical calcula-tions and estimates.

* Arzhannikov@phys.nsu.ru
† I.A.Kotelnikov@inp.nsk.su

None other exact (even in the linear approximation) solutions have been found over the past century. Moreover, even the Havelock solution has not been investigated in details. We do not know any attempt to estimate numerically the integral in his formula for $\zeta$. Various generalizations of the theory of Havelock [9, 54–57] were mainly focused on the computation of the power of wave resistance.

In this paper we propose a new method for solving the problem of ship waves excited by the motion of a submerged object. We demonstrate the prospects of this method by applying it to a moving ball.

In contrast to Havelock, we do not introduce artificial viscosity to ensure the convergence of the integrals. In Sections III and IV, we begin with a solution of the nonstationary problem, suggesting that once in the past a submerged object had been in the state of rest. Transforming our solution to the limit of motion at a constant speed, we automatically arrive at a rule handling the singularity in the integrand, which is completely analogous to the Landau bypass rule in plasma physics [58–60]. The very same singularity in the integrand corresponds to the Cherenkov resonance

$$V\cos\theta = \omega/k, \tag{1}$$

which generates gravity waves with frequency

$$\omega = \sqrt{gk} \tag{2}$$

and wave vector $\mathbf{k}$, which forms an angle $\theta$ with the direction of the velocity. In contrast to the Cherenkov radiation of electromagnetic waves in the optics [61–64], where phase velocity $\omega/k$ has a predetermined value (equal to the speed of light in the medium), due to the dispersion of phase velocity $\omega/k = \sqrt{g/k}$, the gravity waves are emitted in the entire range of angles $\theta$ from 0 (forward in the direction of the body motion) to $\pi$ (against the direction of motion). Each value of the angle $\theta$ corresponds to certain value of the wave number

$$k(\theta) = \frac{g}{V^2\cos^2\theta}. \tag{3}$$

As a consequence, the smallest wave number (i.e., the highest wavelength) that is compatible with the Cherenkov resonance condition is

$$k_g = g/V^2. \tag{4}$$

It should be explained that the Cherenkov resonance concept was not mentioned earlier in the theory of ship waves. Instead, various authors refer to a so called “steady-state condition” or to a “radiation condition”.

In order to compare our method with available literature data, we first attempted to derive a new expression for the vertical elevation $\zeta$ of the liquid surface assuming that a submerged object in the shape of a ball moves with constant velocity parallel to the surface of the liquid. Our expression also contains two terms, as Havelock's solution does. Each term contains single integral and vanishes at $V \to 0$. In the limit of small and large Froude number

$$F = V/\sqrt{gh}, \tag{5}$$

where $h$ denotes the depth of immersion of the ball, we managed to calculate these integrals and obtained relatively simple asymptotic expressions for $\zeta$. The study of these asymptotics shows that the first term can be interpreted as describing the “Bernoulli hump”, and the second term stands for what is called “Kelvin wedge”.

As the second example, we considered purely vertical harmonic oscillatory motion of the ball with a small amplitude. In this case, the solution is expressed in terms of a single integral and describes a radial wave on the liquid surface, diverging from the epicenter over the ball.

In what follows, we adhere the following plan. In Section II, we reproduce, following Ref. [65, §12] in a brief form, the derivation of basic equations that govern gravity waves in order to remind basic assumptions underlying the theory. In Section III, we construct a general solution to the problem of ship waves excited by a source of the pressure acting on the liquid surface. In the next section IV, we show how to transfer this solution to the case of a submerged object that moves along an arbitrary trajectory under the surface of the liquid. In Section V, we proceed to the case of the ball that moves with constant speed parallel to the equilibrium surface of the liquid in order compare our approach with available literature data. In Section V A, we complete the derivation of expression for the elevation of the liquid surface for this case by deforming the path of integration in a complex plane; we named this deformation Peters' transformation after his paper [14]. In Sections V B–V D, we continue analyzing the motion of the ball with constant speed. In Sections V B and V C, we derive approximate formulas for the limiting cases $F \ll 1$ and $F \gg 1$, respectively, and in Section V D we present the results of numerical calculations and discuss qualitative differences in the shape of a liquid surface at various values of the Froude number. In Section VI, we calculate elevation of the surface of the liquid assuming that the submerged ball exhibits small vertical oscillations. Finally, we summarize our findings in Section VII. Appendix A shows how Havelock's result can be derived from our equations.

## II. GRAVITY WAVES ON THE LIQUID SURFACE

Assume that the motion of the fluid can be considered as potential, so its velocity $\mathbf{v}(\mathbf{r}, t)$ at any point, at any time can be expressed in terms of the gradient of potential $\phi(\mathbf{r}, t)$:

$$\mathbf{v} = \nabla\phi. \tag{6}$$

The conditions that justify such approach are detailed in the 6th volume Course of Theoretical Physics [65, §9]. In addition, we assume that the flow velocity is much smaller than the speed of sound in the fluid, so the fluid can be con-sidered as incompressible medium. Then

$$\operatorname{div}\mathbf{v} = 0$$

and the velocity potential satisfies the Laplace equation

$$\nabla^2\phi = 0.$$

The velocity in the potential flow of an incompressible fluid is related to the pressure $p$ and density $\rho$ by equation

$$\frac{\partial\phi}{\partial t} + \frac{1}{2}v^2 + \frac{p}{\rho} + gz = f(t), \tag{7}$$

where $f(t)$ is an arbitrary function of time, and the term $gz$ is added to account for the gravity field. Here and below, we choose a Cartesian coordinate system in which $xy$ plane lies on the equilibrium surface of the liquid, and axis $z$ is directed vertically upward. In the linear approximation, the term $\frac{1}{2}v^2$ in the last equation can be dropped because it contains the square of the speed. The function $f(t)$ can be eliminated by redefining the potential of $\phi$ (adding to $\phi$ a functions of time alone does not change $\mathbf{v}$). However, we will replace the $f(t)$ by the constant $p_0/\rho$, which will be chosen later so as to simplify the subsequent conversion. Then,

$$p = p_0 - \rho g z - \rho\partial\phi/\partial t. \tag{8}$$

Let $\zeta$ be coordinate $z$ of a point on the surface of the liquid; $\zeta$ is a function of three variables: $x$, $y$ and $t$. In equilibrium $\zeta = 0$. When the surface vibrates, $\zeta$ specifies the vertical displacement of the surface.

Suppose that a constant pressure $p_0$ acts on the surface of the liquid. Then, in accordance with the equation (8), we have the relation

$$g\zeta(x,y,t) + \frac{\partial}{\partial t}\phi(x,y,\zeta,t) = 0 \tag{9}$$

on this surface. If vertical displacement $\zeta$ is small compared with the wavelength of oscillation, the vertical component of the velocity of a point is approximately equal to the time derivative of $\zeta$, ie.

$$v_z = \frac{\partial\zeta}{\partial t}. \tag{10}$$

On the other hand, $v_z = \partial\phi/\partial z$, so

$$\frac{\partial}{\partial z}\phi(x,y,\zeta,t) = \frac{\partial}{\partial t}\zeta(x,y,t) = -\frac{1}{g}\frac{\partial^2}{\partial t^2}\phi(x,y,\zeta,t).$$

For oscillations of small amplitude, we can replace the value of the derivatives of potential at $z = \zeta$ in the last equation on their value at $z = 0$. So we arrive at a system of linear partial differential equations derived to describe the motion of a fluid in the gravity field:

$$\nabla^2\phi(x,y,z,t) = 0, \tag{11}$$

$$\frac{\partial}{\partial z}\phi(x,y,0,t) + \frac{1}{g}\frac{\partial^2}{\partial t^2}\phi(x,y,0,t) = 0. \tag{12}$$

We shall here consider waves on the surface of a liquid whose area is unlimited, and we shall also suppose that the wavelength is small in comparison with the depth of the liquid; we can then regard the liquid as infinitely deep. We shall therefore omit the boundary conditions at the sides and bottom (ie., at $z \to -\infty$).

Individual wave on the liquid surface is convenient to characterize by the two-dimensional wave vector

$$\mathbf{k} = (k_x, k_y, 0).$$

A particular solution corresponding to wave with angular frequency $\omega$ can be sought in the form

$$\phi = \Phi(z)\cos(k_x x + k_y y - \omega t).$$

Substituting this expression into Eq. (11), we obtain the equation

$$\frac{\mathrm{d}^2\Phi}{\mathrm{d}z^2} - (k_x^2 + k_y^2)\Phi = 0$$

for the function $\Phi(x)$. Its solution, which decreases as we go into the interior of the liquid (i.e., as $z \to -\infty$) is proportional to $\exp(kz)$ with the exponent

$$k = \sqrt{k_x^2 + k_y^2}, \tag{13}$$

which has the sense of wave number. Consequently,

$$\phi = \phi_k\,\mathrm{e}^{kz}\cos(k_y x + k_y y - \omega t), \tag{14}$$

where $\phi_k$ is the amplitude of the potential, which does not depend on $x$, $y$, $z$, and $t$. The resulting solution must also satisfy the boundary condition (12). Substituting there (14), we obtain the dispersion relation

$$\omega^2 = kg \tag{15}$$

between wave number $k$ and frequency $\omega$ of the gravity wave.

Finally, elevation $\zeta$ of the surface can be found by substituting (14) to Eq. (9), where the second term can be taken again at $z = 0$. This gives

$$\zeta = -\phi_k\frac{k}{\omega}\sin(k_x x + k_y y - \omega t). \tag{16}$$

An arbitrary solution of the system of linear equations (11) and (12) can be constructed as a superposition of particular solutions (14), (16). This will be done in the next Section.

## III. EXCITATION OF SHIP WAVE BY A PRESSURE SOURCE

In this Section, we construct a solution of the Laplace equation (11) with the boundary conditions

$$\frac{\partial}{\partial z}\phi(x,y,0,t) = \frac{\partial}{\partial t}\zeta(x,y,t), \tag{17}$$

$$\frac{\partial}{\partial t}\phi(x,y,0,t) + g\zeta(x,y,t) = -\frac{1}{\rho}\delta p(x,y,t) \tag{18}$$

at the plane $z = 0$, where the pressure is given by the external field $\delta p(x, y, t)$. A general solution of the Laplace equation, decreasing as $z \to -\infty$, has the form

$$\phi(x, y, z, t) = \iint \frac{\mathrm{d}k_x}{2\pi} \frac{\mathrm{d}k_y}{2\pi} \phi_k(t)\, \mathrm{e}^{kz+ik_x x+ik_y y}, \tag{19}$$

where $k = \sqrt{k_x^2 + k_y^2}$. Similarly, we can write the elevation of the liquid surface

$$\zeta(x, y, t) = \iint \frac{\mathrm{d}k_x}{2\pi} \frac{\mathrm{d}k_y}{2\pi} \zeta_k(t)\, \mathrm{e}^{ik_x x+ik_y y} \tag{20}$$

and the external pressure source

$$\delta p(x, y, t) = \iint \frac{\mathrm{d}k_x}{2\pi} \frac{\mathrm{d}k_y}{2\pi} \delta p_k(t)\, \mathrm{e}^{ik_x x+ik_y y} \tag{21}$$

by introducing the amplitude functions of time $\zeta_k(t)$ and $\delta p_k(t)$. For these functions we obtain the equation

$$k\phi_k(t) = \frac{\partial}{\partial t}\zeta_k(t),$$
$$\frac{\partial}{\partial t}\phi_k(t) + g\zeta_k(t) = -\frac{1}{\rho}\delta p_k(t)$$

from the boundary conditions. Excluding $\phi_k(t)$ from these equations, we obtain the ordinary differential equation of the second order for the function $\zeta_k(t)$:

$$\frac{\partial^2}{\partial t^2}\zeta_k(t) + kg\zeta_k(t) = -\frac{k}{\rho}\delta p_k(t). \tag{22}$$

In addition, we assume that in the distant past, there was no external pressure source and, respectively, the liquid surface was quite flat. With this formulation of the problem, one has to consider that $\zeta_k(t) \to 0$ and $\partial\zeta_k(t)/\partial t \to 0$ as $t \to -\infty$. Corresponding solution of Eq. (22) has the form

$$\zeta_k(t) = \frac{\sqrt{kg}}{\rho g} \int_{-\infty}^{t} \sin\left[\sqrt{kg}(\tau - t)\right] \delta p_k(\tau)\, \mathrm{d}\tau. \tag{23}$$

Finally, we take into account that

$$\delta p_k(t) = \iint \mathrm{d}x\, \mathrm{d}y\, \delta p(x, y, t)\, \mathrm{e}^{-ik_x x-ik_y y}. \tag{24}$$

Convergence of the integral in Eq. (23) for realistic functions $\delta p(x, y, t)$ is guaranteed by the fact that $\delta p(x, y, t)$ tends to zero at $t \to -\infty$. To pass to the case of motion at a constant speed, while maintaining the convergence of the integral (23) at lower limit $t \to -\infty$, we choose function $\delta p(x, y, t)$ in an appropriate form by adding factor $\exp(\mu t)$ with a parameter $\mu > 0$:

$$\delta p(x, y, t) = \mathrm{e}^{\mu t}\, \delta\hat{p}(x - Vt, y); \tag{25}$$

later, we will take the limit $\mu \to 0$. Putting function (25) in the integral (24), we make the substitution $x \to x + Vt$, which corresponds to the transition to moving reference frame of the pressure source. We then obtain

$$\delta p_k(t) = \mathrm{e}^{\mu t-ik_x Vt}\, \delta\hat{p}_k, \tag{26}$$

where

$$\delta\hat{p}_k = \iint \mathrm{d}x\, \mathrm{d}y\, \delta\hat{p}(x, y)\, \mathrm{e}^{-ik_x x-ik_y y}. \tag{27}$$

After substituting (26) in Eq. (23) the integration over time yields

$$\zeta_k(t) = \frac{k\delta\hat{p}_k}{((k_x V + i\mu)^2 - kg)\,\rho}\, \mathrm{e}^{\mu t-ik_x Vt}. \tag{28}$$

Pitting this result in Eq. (20) we find that in the fixed reference frame

$$\zeta(x, y, t) = \frac{1}{\rho} \iint \frac{\mathrm{d}k_x}{2\pi} \frac{\mathrm{d}k_y}{2\pi} \frac{k\, \delta\hat{p}_k\, \mathrm{e}^{\mu t+ik_x(x-Vt)+ik_y y}}{(k_x V + i\mu)^2 - kg}.$$

In the frame of reference moving with the source of pressure, in the limit $\mu \to 0$ we obtain stationary elevation of the liquid surface

$$\zeta(x, y, t) = \frac{1}{\rho} \iint \frac{\mathrm{d}k_x}{2\pi} \frac{\mathrm{d}k_y}{2\pi} \frac{k\, \delta\hat{p}_k\, \mathrm{e}^{ik_x x+ik_y y}}{(k_x V + i0)^2 - kg}. \tag{29}$$

This result coincides with Eq. (2.17b) in [54], considering that the latter was written for the motion in negative direction of the axis of $x$ and therefore $V$ should be changed to $-V$.

The addition of $i0$ in the denominator of the integrand in Eq. (29) symbolizes that the singularity of the integrand is in fact shifted from the path of integration to the complex plane. Thus, $i0$ in the denominator gives a bypass rule similar to the Landau bypass rule in the theory of plasma oscillations. We get it right quite in the same manner as it has been derived by Lev Landau [58, 59]. Namely, we investigated the excitation of gravity waves by solving the initial-value problem starting with an instant of time in the past when the pressure source had been in the state of rest. Parameter $\mu > 0$ was introduced only in order to make the passage to the limit of motion at a constant speed.

Interestingly, that Horace Lamb, faced with the problem of the divergence of the integral because of the singularity on the contour of integration [10, 11], tried to interpret the integral in the sense of the principal value without any explanation, but added a term, which, in the sum with principal value of the integral, yielded zero elevation of the liquid surface far ahead of the external pressure source. In this way, he got a correct result, but missed an opportunity to discover the Landau bypass rule, which has application in various fields of physics (see., eg, [60, 66]). In the above mentioned paper by Raphaël [54], an imaginary component in the denominator (29) was added with reference to Lighthill's monograph [5, §3.9], and James Lighthill, in his turn, appealed to the "radiation condition". Our derivation in this Section provides a rigorous justification to all these intuitive approaches.

## IV. EXCITATION OF SHIP WAVE BY A SUBMERGED BALL

Suppose that a ball of radius $a$ is moving below the equilibrium surface of the liquid at a depth $h \gg a$. Here again, we shall use Cartesian system of coordinate $x$, $y$, and $z$, introduced in the previous Section. We assume that the ball is moving with variable velocity $\mathbf{V} = (\dot{X}(t), \dot{Y}(t), \dot{Z}(t))$, and functions $X(t)$, $Y(t)$, $Z(t)$ determine the coordinates of the ball center at any instant of time $t$.

Given that $h = -Z(t) \gg a$, we seek a solution of the Laplace equation (11) as the sum

$$\phi = \phi_0 + \phi_1 \tag{30}$$

of potential $\phi_0$, which is the solution of the problem of potential motion of a body in an unbounded liquid [65, §11], and an additive $\phi_1$, which is necessary to satisfy the boundary conditions on the surface of the liquid. Total potential $\phi$ and elevation of the surface $\zeta$ satisfy boundary conditions (17) and (18), in which one needs to drop external pressure $\delta p$:

$$\frac{\partial}{\partial z}\phi(x,y,0,t) = \frac{\partial}{\partial t}\zeta(x,y,t), \tag{31}$$

$$\frac{\partial}{\partial t}\phi(x,y,0,t) + g\zeta(x,y,t) = 0. \tag{32}$$

The first part of potential

$$\phi_0(x,y,z,t) = = -\frac{\partial}{\partial t}\frac{a^3/2}{\sqrt{(x-X(t))^2+(y-Y(t))^2+(z-Z(t))^2}} \tag{33}$$

at $z + Z(t) > 0$ can be represented as

$$\phi_0(x,y,z,t) = \iint \frac{\mathrm{d}k_x}{2\pi}\frac{\mathrm{d}k_y}{2\pi}\, \mathrm{e}^{-kz+ik_x x+ik_y y}\, \phi_k^{(0)}(t), \tag{34}$$

where

$$\phi_k^{(0)}(t) = -\frac{\pi a^3}{k}\frac{\partial}{\partial t}\, \mathrm{e}^{kZ(t)-ik_xX(t)-ik_yY(t)}. \tag{35}$$

In addition, we use the assumption that in the distant past the ball had been in the state of rest before it began to move and, therefore, $\phi_k^{(0)}(-\infty) = 0$. It is important that, because of the multiplier $\mathrm{e}^{-kz}$, the integrand in Eq. (34) grows as we go deeper into the liquid. On the contrary, for $\phi_1$, we need to take a solution that decreases with $z \to -\infty$, as is done in Section III. Therefore, by analogy with Eq. (19) we can write

$$\phi_1(x,y,z,t) = \iint \frac{\mathrm{d}k_x}{2\pi}\frac{\mathrm{d}k_y}{2\pi}\, \mathrm{e}^{kz+ik_x x+ik_y y}\, \phi_k^{(1)}(t). \tag{36}$$

Substituting now Eqs. (20), (34) and (36) in boundary conditions (31), (32) and excluding $\phi_k^{(1)}$, we obtain the equation

$$\frac{\partial^2}{\partial t^2}\zeta_k(t) + kg\zeta_k(t) = -2k\frac{\partial}{\partial t}\phi_k^{(0)}(t) \tag{37}$$

for function $\zeta_k(t)$. Since it differs from Eq. (22) by the right-hand-side only, all the results, obtained in Section III, are transferred to the current case by replacing $\delta p_k/\rho \to 2\partial\phi_k/\partial t$. Making this change in Eq. (23) and then integrating by parts, taking into account the conditions $\phi_k^{(0)}(-\infty) = 0$, we find that

$$\zeta_k(t) = -2k\int_{-\infty}^{t} \cos\left[\sqrt{kg}(\tau - t)\right]\phi_k^{(0)}(\tau)\,\mathrm{d}\tau. \tag{38}$$

Substituting (35) in Eq. (38), we obtain a formula for the Fourier amplitude of the vertical elevation of the liquid surface, which is created by a submerged ball, moving in an arbitrary way:

$$\zeta_k(t) = -2k\int_{-\infty}^{t} \cos\left[\sqrt{kg}(\tau - t)\right]\phi_k^{(0)}(\tau)\,\mathrm{d}\tau = = 2\pi a^3 \int_{-\infty}^{t} \cos\left[\sqrt{kg}(\tau - t)\right]\frac{\partial}{\partial \tau}\,\mathrm{e}^{kZ(\tau)-ik_xX(\tau)-ik_yY(\tau)}\,\mathrm{d}\tau. \tag{39}$$

Having found the Fourier-amplitude $\zeta_k(t)$, one can restore function $\zeta(x,y,t)$ using Eq. (20), although computation of the involved integrals represents a challenge task.

Convergence in the integral (39) for more or less realistic functions $X(t)$, $Y(t)$, $Z(t)$ is guaranteed by the fact that their time derivatives tend to zero as $t \to -\infty$.

It is worth noting that Eq. (38) can be used to construct a solution for a submerged object other than a ball if we put corresponding function $\phi_k^{(0)}(\tau)$. For example, this function can be easily found for a prolate ellipsoid. Main difficulty in this case is transferred to the computation of the integrals in Eqs. (38) and (20).

## V. UNIFORM MOTION OF THE BALL

To proceed to the limit of motion of the ball with a constant velocity $V = \mathrm{const}$, parallel to the surface of the equilibrium liquid, we assume that

$$X(t) = Vt, \qquad Y(t) = 0, \qquad Z(t) = -h. \tag{40}$$

However, to keep convergence of integral (39) at the lower limit $t = -\infty$, we add factor $\mathrm{e}^{\mu\tau}$ with $\mu > 0$ to the integrand in order to emulate the condition $\dot{X}(-\infty) = 0$ used in the derivation of the integral. Then, the integration over time can be performed in general form and the result is

$$\zeta_k = 2\pi a^3\,\mathrm{e}^{-kh-ik_xVt+\mu t}\,\frac{k_xV\,(k_xV+i\mu)}{(k_xV+i\mu)^2 - kg}. \tag{41}$$

Seeking the limit $\mu \to 0$ of this expression, one needs to keep a rudiment of $\mu$ only in the denominator because the denominator vanishes at the Cherenkov resonance

$$kg = k_x^2V^2. \tag{42}$$

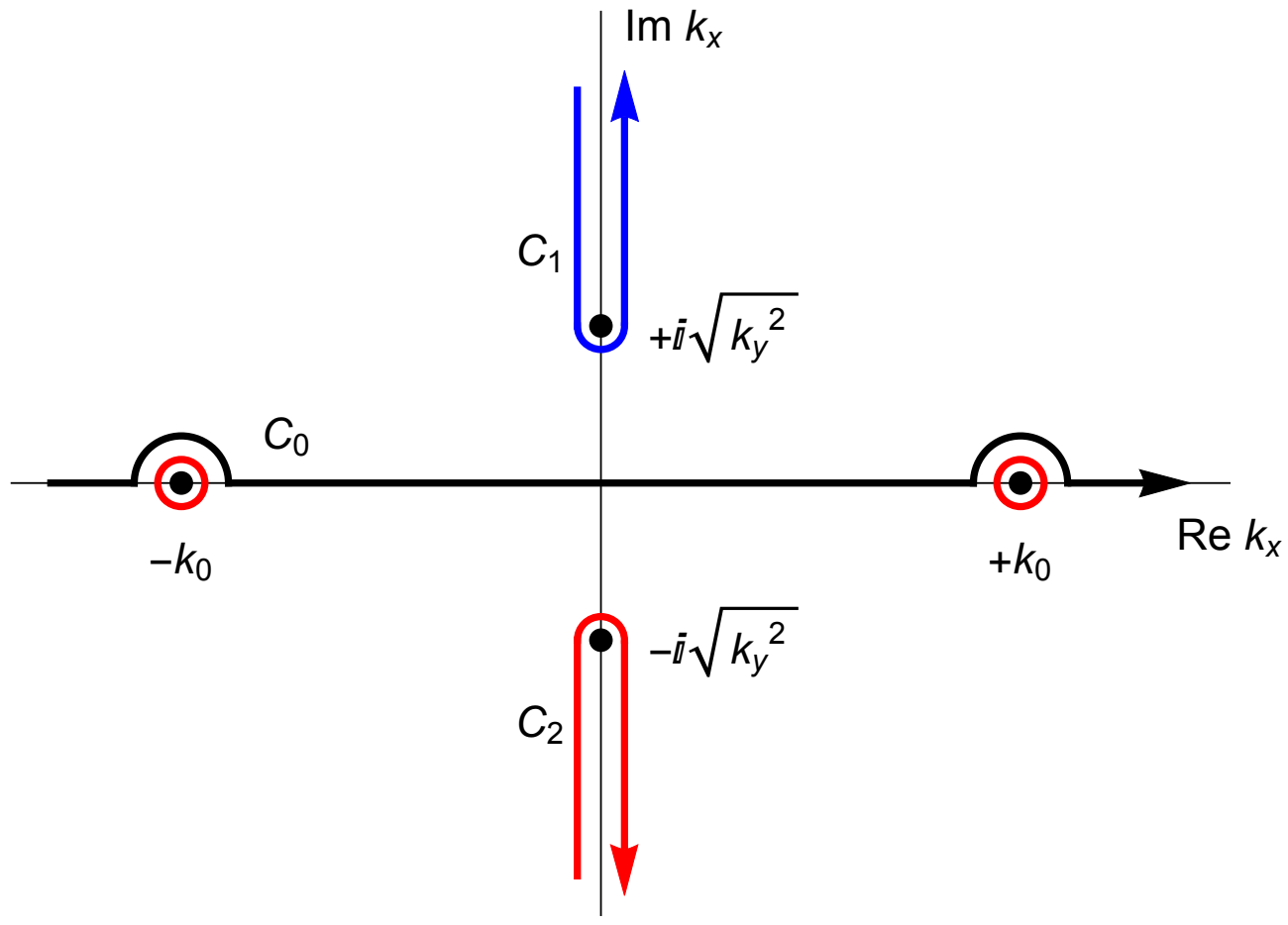

Figure 1. (Color online) The contour of integration in the complex plane $k_x$ at $V > 0$: original contour of integration $C_0$ (black curve) circumvents above the poles of the integrand at the points $k_x = \pm k_0$; if $x > 0$ deformed contour $C_1$ (blue curve) in the upper half-plane comes from infinity $+i\infty$ on the left side of the branch-cut $[i\sqrt{k_y^2}, i\infty)$, bypasses the branch point $k_x = +i\sqrt{k_y^2}$ from below and goes to infinity on the right side of the cut; if $x < 0$ the deformed contour (red line) goes around the poles at $k_x = \pm k_0$ in a clockwise direction, the other part $C_2$ of the deformed contour passes along the branch-cut $[-i\sqrt{k_y^2}, -i\infty)$, bypassing the branch point $k_x = -i\sqrt{k_y^2}$ at the lower half-plane.

Putting (41) in Eq. (20), we also perform transformation to the reference frame of the moving ball by making the substitution $x \to x + Vt$. This yields the double integral

$$\zeta(x,y) = \frac{a^3}{2\pi} \iint \mathrm{d}k_x\, \mathrm{d}k_y \frac{k_x^2 V^2\, \mathrm{e}^{-kh+ik_x x+ik_y y}}{(k_x V + i0)^2 - kg}, \tag{43}$$

which depends on the coordinates $x, y$ as external parameters. The term $i0$ (ie, $i\mu$ in the limit $\mu \to 0$) in the denominator of the integrand in Eq. (43) gives a crawl rule around the singularity similar to the already mentioned Landau bypass rule in the theory of plasma oscillations [60].

The exponential factor $\mathrm{e}^{-kh}$ in the integrand in Eq. (43) makes a natural scale of the wave number $k_h \sim 1/h$. Another scale $k_g \sim g/V^2$ is defined by the Cherenkov resonance. One should therefore expect that the wavelengths $\lambda_h \sim h$ and $\lambda_g \sim V^2/g$ will dominate in function $\zeta(x,y)$.

Concluding this section, we note that Havelock's formula for the surface elevation [7, 8] can be obtained by making one more integration by parts in Eq. (39). This transformation is done in Appendix A. In the next Section, we calculate the integral in Eq. (43) in another way, which leads to the expression, the component parts of which have a simple physical meaning, in contrast to those in Havelock's formula.

### A. Peters' transformation

Proceeding to calculation of the integral (43), first we integrate over the variable $k_x$:

$$\zeta(x,y) = \frac{a^3}{2\pi} \int_{-\infty}^{\infty} \mathrm{d}k_y\, \mathrm{e}^{ik_y y} \int_{-\infty}^{\infty} \mathrm{d}k_x \frac{k_x^2 V^2\, \mathrm{e}^{-kh+ik_x x}}{(k_x V + i0)^2 - kg}. \tag{44}$$

The integrand in

$$\int_{C_0} \mathrm{d}k_x \frac{k_x^2 V^2\, \mathrm{e}^{-kh+ik_x x}}{(k_x V + i0)^2 - kg} \tag{45}$$

has first-order poles at the points

$$k_x = \pm k_0, \qquad k_0 = \frac{g}{\sqrt{2}V^2}\sqrt{1 + \sqrt{1 + 4k_y^2 V^4/g^2}}, \tag{46}$$

where the denominator of the integrand vanishes. Due to the imaginary additive $i0$ in the denominator, the poles are shifted from the real axis $\mathrm{Re}\, k_x$. They are shifted to the lower half of the complex plane $k_x = \mathrm{Re}\, k_x + i\, \mathrm{Im}\, k_x$, if $V > 0$, and to its upper half, if $V < 0$. This means that the integration contour $C_0$ in Eq. (45) passes above these points if $V > 0$, and below if $V < 0$. We dwell on the case of $V > 0$ for which integration contour $C_0$ is shown in Fig. 1.

To transform the integral (45) to a form more suitable for calculations, we deform original contour $C_0$ by moving it to upwards or downwards in the complex plane $k_x$. The deformed contour cannot cross singularities of the integrand (otherwise the result of integration would change) and therefore it "catches" on the singularities. In addition to the first-order pole $k_x = \pm k_0$, the integrand in Eq. (45) has two branch points $k_x = \pm i\sqrt{k_y^2}$, where $k = 0$. Choosing the branch of the square root of $k = \sqrt{k_x^2 + k_y^2}$, which is positive for real $k_x, k_y$, we make a branch-cut along the imaginary axis from the branch points $k_x = +i\sqrt{k_y^2}$ in the upper half-plane $\mathrm{Im}\, k_x > 0$ up to $k_x = +i\infty$, and another branch-cut in the lower half-plane $\mathrm{Im}\, k_x < 0$ from the branch point $k_x = -i\sqrt{k_y^2}$ downwards.

According to Jordan's lemma [67, p. 272], the integral over the semicircle of infinite radius in the upper half becomes zero if $x > 0$, while the integral over infinite semicircle in the lower half vanishes if $x < 0$.

Complementing the original path of integration $C_0$ by such semicircles and using Cauchy's residue theorem [67, p. 234] is not difficult to prove that the integral over $C_0$ for $x > 0$ is equal to the integral over contour $C_1$, which goes along the edges of the branch-cut in the upper half-plane, as shown in Fig. 1. Indeed, joining the contours $C_0$ and $C_1$ with arcs of the infinite semicircle, we obtain a closed circuit. The integral around such closed contour vanishes since it has no poles inside. On the other hand, it is equal to the difference between the integrals over the contours $C_0$ and $C_1$ as the integral over the infinite semicircle is zero, and contour $C_1$ is included into the closed contour in the direction opposite to that of contour $C_0$. Consequently,

$$\int_{C_0} \mathrm{d}k_x(\ldots) - \int_{C_1} \mathrm{d}k_x(\ldots) = 0, \qquad (47)$$

if $x > 0$. In case $x < 0$, in the similar way we can prove that the difference of the integrals over the contours $C_0$ and $C_2$ is equal to the sum of the residues at the poles of $k_x = \pm k_0$, multiplied by $-2\pi i$, as these poles are encircled by the closed contour, which bypasses them in the negative direction (clockwise). Hence,

$$\int_{C_0} \mathrm{d}k_x(\ldots) - \int_{C_2} \mathrm{d}k_x(\ldots) = -2\pi i \sum_{k_x=\pm k_0} \mathrm{Res}(\ldots). \qquad (48)$$

The integral along contour $C_1$ for $x > 0$ can be written as the sum of the integrals on the left and right edges of the branch-cut in the upper half-plane:

$$\int_{C_1} \mathrm{d}k_x(\ldots) = \int_{i\infty-0}^{i\sqrt{k_y^2}-0} \mathrm{d}k_x\,(\ldots) + \int_{i\sqrt{k_y^2}+0}^{i\infty+0} \mathrm{d}k_x\,(\ldots)\,.$$

In the integral on the left edge (the first term) we make the substitution $k_x = i\sqrt{\varkappa^2 + k_y^2} - 0$, with $k = \sqrt{k_x^2 + k_y^2} = -i\sqrt{\varkappa^2} = i\varkappa$, if we assume that $\varkappa < 0$ on this edge. In the integral on the right edge (the second term) we make the substitution $k_x = i\sqrt{\varkappa^2 + k_y^2} + 0$, with $k = +i\sqrt{\varkappa^2} = i\varkappa$, if $\varkappa > 0$. Thus, the first term is converted to an integral over $\varkappa$ from $-\infty$ to 0, and the second from 0 to $+\infty$. Considering also that the $k_x\,\mathrm{d}k_x = -\varkappa\,\mathrm{d}\varkappa$, we have

$$\int_{C_1} \mathrm{d}k_x(\ldots) = \int_{-\infty}^{\infty} \mathrm{d}\varkappa\, \frac{i\varkappa\sqrt{\varkappa^2 + k_y^2}\, \mathrm{e}^{-i\varkappa h - \sqrt{\varkappa^2 + k_y^2}\, x}}{\varkappa^2 + k_y^2 + i\varkappa g/V^2}. \qquad (49)$$

Similarly, the integral over $C_2$ for $x < 0$ can be written as the sum of the integrals on the left and right edges of the cut in the lower half-plane:

$$\int_{C_2} \mathrm{d}k_x(\ldots) = \int_{-i\infty-0}^{-i\sqrt{k_y^2}-0} \mathrm{d}k_x\,(\ldots) + \int_{-i\sqrt{k_y^2}+0}^{-i\infty+0} \mathrm{d}k_x\,(\ldots)\,.$$

In the integral on the left edge (the first term) we make the substitution $k_x = -i\sqrt{\varkappa^2 + k_y^2} - 0$, with $k = \sqrt{k_x^2 + k_y^2} = +i\sqrt{\varkappa^2} = i\varkappa$, if we assume that $\varkappa > 0$. In the integral on the right edge (the second term) we make the substitution $k_x = -i\sqrt{\varkappa^2 + k_y^2} + 0$, with $k = -i\sqrt{\varkappa^2} = i\varkappa$, if $\varkappa < 0$. Thus, the first term is converted to the integral over $\varkappa$ going from $+\infty$ and 0, and the second term to the integral from 0 to $-\infty$. The resulting expression differs from (49) by replacing $x$ to $-x$. Combining both integrals over $C_1$ and $C_2$ and restoring integration over $k_y$, we introduce the function

$$\zeta_0(x,y) = \frac{a^3}{2\pi} \int_{-\infty}^{\infty} \mathrm{d}k_y\, \mathrm{e}^{ik_y y} \times$$
$$\times \int_{-\infty}^{\infty} \mathrm{d}\varkappa \frac{i\varkappa\sqrt{\varkappa^2 + k_y^2}\, \mathrm{e}^{-i\varkappa h - \sqrt{\varkappa^2 + k_y^2}\,|x|}}{\varkappa^2 + k_y^2 + i\varkappa g/V^2}. \qquad (50)$$

Contribution of the residues in Eq. (48) generates the function

$$\zeta_1(x,y) = \frac{a^3}{2\pi} \int_{-\infty}^{\infty} \mathrm{d}k_y\, \mathrm{e}^{ik_y y} \left( -2\pi i \sum_{k_x=\pm k_0} \mathrm{Res}(\ldots) \right),$$

so that the total solution for $x < 0$ (we continue to analyze the case of $V > 0$) is

$$\zeta(x,y) = \zeta_0(x,y) + \zeta_1(x,y). \qquad (51)$$

The residue at $k_x = \pm k_0$ is equal to the limit $k_x \to \pm k_0$ of the integrand multiplied by $k_x - (\pm k_0)$. As a result of simple calculations, we find that

$$\zeta_1 = -\frac{2a^3V^2}{g} \int_{-\infty}^{\infty} \mathrm{d}k_y \frac{k_* k_0\, \mathrm{e}^{ik_y y - k_* h}}{\left(1 + 4V^4 k_y^2/g^2\right)^{1/2}} \sin\left(k_0|x|\right) \qquad (52)$$

where

$$k_* = \sqrt{k_0^2 + k_y^2} = \frac{g}{2V^2}\left[1 + \sqrt{1 + 4V^4 k_y^2/g^2}\right] \qquad (53)$$

and we have taken into account that $x = -|x|$.

Transition to the case of $V < 0$ is quite simple. If you change the sign of the velocity, it is just enough to change the sign of coordinate $x$. Since functions $\zeta_0$ and $\zeta_1$ are defined so that they do not depend on the speed sign, the result of our calculations can be written in the following universal form, suitable for any sign of $V$:

$$\zeta(x,y) = \zeta_0(|x|,y) + \zeta_1(|x|,y)H(-Vx), \qquad (54)$$

where $H(z)$ is the Heaviside function, which is 1 for $x > 0$ and 0 for $x < 0$. We also note that functions $\zeta_0$ and $\zeta_1$ are real, although the integrands in their definition are complex. This follows from the fact that the complex conjugate of the integrands in Eqs. (50) and (52) is equivalent to reversing the sign of $k_y$.

It is useful to present Eq. (50) in the form

$$\zeta_0(x,y) = \frac{a^3}{2\pi} \int_{-\pi}^{\pi} \mathrm{d}\psi \int_0^{\infty} \mathrm{d}q\, \mathrm{e}^{iqy\cos\psi - iqh\sin\psi - q|x|} \times$$
$$\times \frac{iq^2 \sin\psi}{q + i(g/V^2)\sin\psi} \qquad (55)$$

by making the change of variables

$$k_y = q\cos\psi, \qquad \varkappa = q\sin\psi. \tag{56}$$

It has no obvious physical sense, but allows to perform integration over variable $q$, which yields

$$\zeta_0(x,y) = \frac{ia^3g^2}{2\pi V^4}\int_{-\pi}^{\pi} \mathrm{d}\psi\,\frac{\sin\psi}{R^2}\{1 - iR\sin\psi + \\ + \mathrm{e}^{iR\sin\psi}R^2\sin^2\psi\,[\mathrm{Ci}(R\sin\psi) - i\,\mathrm{Si}(R\sin\psi) \\ + \ln(R) - \ln(-i/\sin\psi) - \ln(R\sin\psi)]\}, \tag{57}$$

where

$$R = (g/V^2)\,(|x| - iy\cos\psi + ih\sin\psi),$$
$$\mathrm{Ci}(z) = -\int_z^{\infty}\cos(t)/t\ \mathrm{d}t,$$
$$\mathrm{Si}(z) = \int_0^z \sin(t)/t\ \mathrm{d}t.$$

The functions of logarithm $\ln(z)$ and integral cosine $\mathrm{Ci}(z)$ (in contrast to the integral sine $\mathrm{Si}(z)$) have gaps at the edges of the cut in the complex plane $z$ going from 0 to $-\infty$, but the total integrand in (57) is everywhere continuous.

The integrand in (57) in the limit $x \to 0$ is singular at two points on the interval $\psi \in [-\pi,\pi]$, where $\mathrm{tg}\,\psi = y/h$. This feature is the consequence of the divergence of integral over $q$ in Eq. (55) at $x = 0$.

Finally, we note that the method of converting double integral (44), described in this section, was inspired by A. Peters' paper [14]. We have advanced his technique by reducing the number of changes of integration variables.

### B. Limit of small Froude number

In the limit of small Froude number $F = V/\sqrt{gh}$, $F \ll 1$, functions $\zeta_0(x,y)$ and $\zeta_1(x,y)$ in Eq. (54) can be calculated in closed form. This case can be also called the limit of low speed. If $V \to 0$, it is sufficient to keep only the last term in the denominator of the integrand in Eq. (55). This leads to the following expression for $\zeta_0(x,y)$:

$$\zeta_0 = \frac{a^3V^2}{2\pi g}\int_{-\pi}^{\pi}\mathrm{d}\psi\int_0^{\infty}\mathrm{d}q\,q^2\,\mathrm{e}^{iqy\cos\psi - iqh\sin\psi - q|x|} = \\ = \frac{a^3V^2}{2\pi g}\frac{\partial^2}{\partial x^2}\int_{-\pi}^{\pi}\mathrm{d}\psi\int_0^{\infty}\mathrm{d}q\,\mathrm{e}^{iq\sqrt{h^2+y^2}\cos\psi - q|x|}.$$

The result of calculations

$$\zeta_0 = -\frac{a^3V^2}{g}\frac{h^2 - 2x^2 + y^2}{(x^2+y^2+h^2)^{5/2}} \tag{58}$$

is exactly 2 times greater than the elevation of the liquid surface evaluated with the aid of Eq. (32) if we would drop $\phi_1$ term in Eq. (30) and keep only the potential

$$\phi_0(x,y,z) = \frac{\partial}{\partial x}\frac{a^3V/2}{\sqrt{x^2+y^2+(z+h)^2}}$$

of the ball in unbounded liquid; note that the transition to the ball reference frame is performed by replacing $\partial/\partial t$ in Eq. (32) with $V\partial/\partial x$. The doubling of the elevation has a simple explanation. Indeed, if we take

$$\phi_1(x,y,z) = \phi_0(x,y,-z), \tag{59}$$

the sum $\phi = \phi_0 + \phi_1$ will automatically satisfy boundary condition

$$\frac{\partial\phi}{\partial z} = 0, \tag{60}$$

which represent ultimate form of Eq. (31) at $V \to 0$, as the right-hand side of Eq. (31) is proportional to $V$. Substituting $\phi_0(x,y,z) + \phi_0(x,y,-z)$ in Eq. (32) then leads to Eq. (58).

Formula (58) describes a recess on the liquid surface located above the center of the ball. Its depth

$$\zeta_0 \sim \frac{a^3}{h^2}\frac{V^2}{gh} = \frac{a^3}{h^2}F^2 \tag{61}$$

can be used as an estimate of the liquid surface displacement amplitude in the case $F \ll 1$.

To calculate the second term $\zeta_1(x,y)$ in Eq. (54) in the case of small velocity, we substitute

$$k_* \approx g/V^2 + k_y^2V^2/g, \qquad k_0 \approx g/V^2 + k_y^2V^2/2g \tag{62}$$

in Eq. (52) in those terms where $k_*$ and $k_0$ enter the arguments of the exponential and trigonometric functions, and drop $V^2$ terms in Eq. (62) in other cases. The thus obtained integral

$$\zeta_1 = -\frac{2a^3g}{V^2}\int_{-\infty}^{\infty}\mathrm{d}k_y\,\mathrm{e}^{ik_yy - hk_y^2V^2/g - gh/V^2}\times \\ \times\sin\left(k_y^2V^2|x|/2g + g|x|/V^2\right) \tag{63}$$

was calculated using Wolfram Mathematica [68]:

$$\zeta_1 = \frac{2\sqrt{2\pi}a^3g^{3/2}}{|V|^3}\exp\left(-\frac{gh}{V^2}\frac{4h^2+x^2+y^2}{4h^2+x^2}\right)\times \\ \times\,\mathrm{Im}\left[\frac{1}{\sqrt{2h+i|x|}}\exp\left(-i\frac{g|x|}{2V^2}\frac{8h^2+2x^2-y^2}{06h^2+x^2}\right)\right]. \tag{64}$$

Using this result, we can estimate the amplitude of the first (deepest) depressions on the profile of $\zeta_1(x,y)$ as

$$\min\zeta_1 = -2\sqrt{\pi}\frac{a^3g^{3/2}}{|V|^3\sqrt{h}}\exp\left(-\frac{gh}{V^2}\right). \tag{65}$$

Comparing it with the value of the expression (58), we conclude that $|\zeta_1| \ll |\zeta_0|$ in the case of small Froude number.

Formula (64) provides a reasonable accuracy even if $F \sim 1$. This fact is illustrated in Fig. 2, where profiles of $\zeta(x,y)$ for $F = 0.5$ are plotted using both exact and approximate formulas. It is worth noting that approximate computation with Wolfram Mathematica 10 [68] was 1500 times faster.

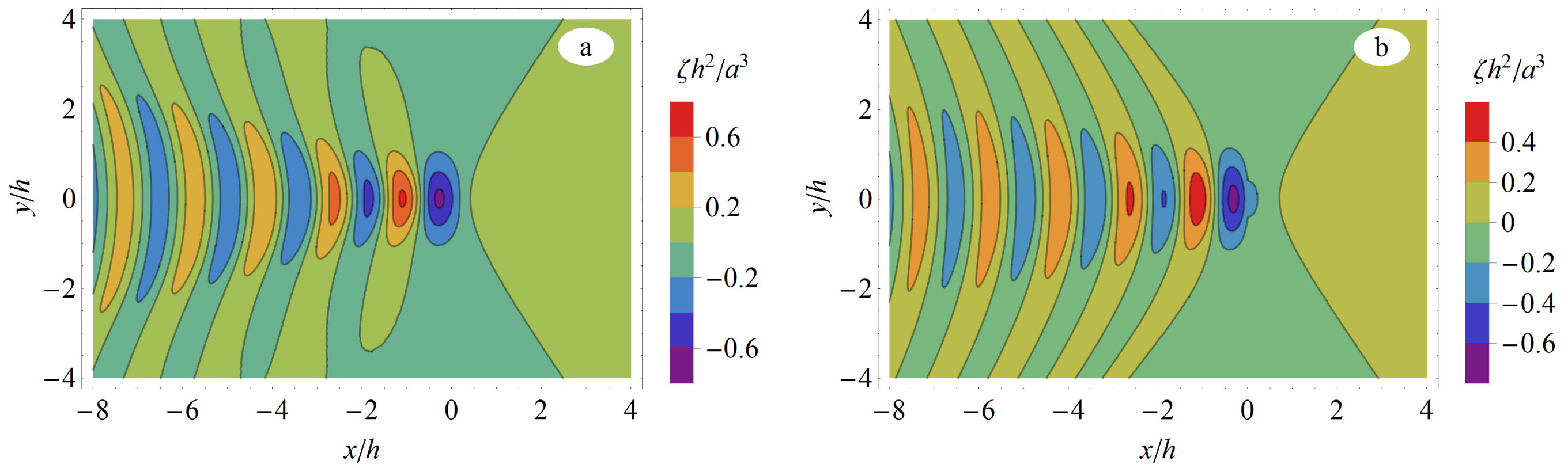

Figure 2. (Color online) Isolines of function $(a^3/h^2)^{-1}\zeta(x,y)$ at $F = 0.5$: (a) exact calculation using Eqs. (50) and (52), (b) approximate calculation using Eqs. (58) and (64).

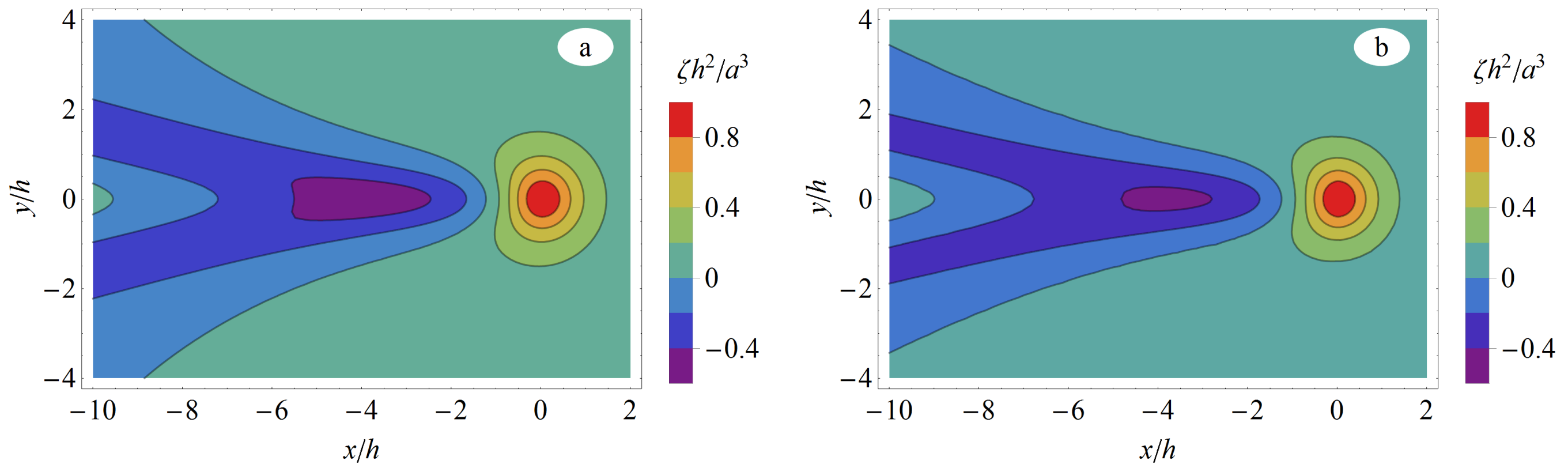

Figure 3. (Color online) Isolines of function $(a^3/h^2)^{-1}\zeta(x,y)$ at $F = 3$: (a) accurate calculation using Eqs. (50) and (52), (b) approximate calculation using Eqs. (66) and (68).

### C. Limit of large Froude number

In the limit of large Froude number, $F \gg 1$, a principal term in $\zeta_0$ does not depend on the ball speed. Keeping only the first term in the denominator of the integrand in Eq. (55) for $\zeta_0(x,y)$, we obtain

$$\zeta_0 = \frac{a^3}{2\pi}\int_{-\pi}^{\pi} \mathrm{d}\psi \int_0^{\infty} \mathrm{d}q\, iq \sin\psi \times \\ \times \mathrm{e}^{iqy\cos\psi - iqh\sin\psi - q|x|} = \frac{a^3 h}{(x^2+y^2+h^2)^{3/2}}. \quad (66)$$

In this case, function $\zeta_0(x,y)$ describes a hump on the surface of the liquid, which is traditionally called the Bernoulli hump. Its height

$$\zeta_0 = \frac{a^3}{h^2} \quad (67)$$

can be used as an estimate of the liquid surface displacement amplitude in the case $F \gg 1$.

To calculate integral (52) in the high-velocity limit, we substitute

$$k_* \approx \sqrt{k_y^2 + \frac{g}{2V^2}}, \qquad k_0 \approx \sqrt[4]{\frac{k_y^2 g^2}{V^4}}.$$

The resulting integral

$$\zeta_1 = -a^3 \int_{-\infty}^{\infty} \mathrm{d}k_y \sqrt[4]{\frac{k_y^2 g^2}{V^4}} \times \\ \times \mathrm{e}^{ik_y y - \sqrt{k_y^2}h - gh/2V^2} \sin\left(\sqrt[4]{\frac{k_y^2 g^2}{V^4}}|x|\right)$$

was again computed by Wolfram Mathematica [68]:

$$\zeta_1(x,y) = \frac{a^3\sqrt{g}}{4V^3}\,\mathrm{e}^{-gh/2V^2}\left\{\frac{4\sqrt{g}x\,|V|\left(h^2-y^2\right)}{\left(h^2+y^2\right)^2} + \right. \\ +2\sqrt{\pi}\,\mathrm{Re}\left[\frac{2hV^2 - gx^2 - 2iV^2 y}{(h-iy)^{5/2}} \times \right. \\ \left.\left. \times \exp\left(-\frac{gx^2}{4V^2(h-iy)}\right)\mathrm{erfi}\left(\frac{\sqrt{g}x}{2\,|V|\sqrt{h-iy}}\right)\right]\right\}, \quad (68)$$

where $\operatorname{erfi}(z) = \operatorname{erf}(iz)/i$ is the complex error function. Minimum of function (68) is achieved at distance $x \approx -1.19\,Fh$ at the $x$ axis behind the ball, where a depression is formed on the liquid surface, following the Bernoulli hump (66). Its depth

$$\zeta_1 = -1.47\frac{a^3}{Fh^2} \tag{69}$$

can be used to estimate the value of $\zeta_1$ in order of magnitude. Here again it turns out that $|\zeta_1| \ll |\zeta_0|$, as in the case $F \ll 1$, if one compares the amplitudes of functions $\zeta_1$ and $\zeta_0$. However, as will be seen in Fig. 5 in the next section, on the edges of the Kelvin wedge, on the contrary, $|\zeta_1| \gg |\zeta_0|$ if $F \gg 1$. Furthermore, the amplitudes of $\zeta_1$ and $\zeta_0$ have the same order of magnitude at $F \sim 1$.

In Fig. 3 the results of computation of function $\zeta(x, y)$ using the exact and approximate formulas are shown for $F = 3$; approximate calculation with Wolfram Mathematica 10 [68] was 500 times faster.

### D. Profile of the ship wave

To demonstrate qualitative changes in the shape of the liquid surface, which occur with increasing Froude number, we have drawn the profile of elevation $\zeta(x, 0)$ of the liquid surface in $xz$ plane, passing through the center of the ball. The results of calculations by the exact formulas (52), (54) and (57) are shown in Fig. 4 for the six values of the Froude number. In the same figures, dashed lines show the same profiles computed by the approximate formulas (58) and (64), intended for the case $F \ll 1$, and dotted lines are drawn by the approximate formulas (66) and (68), derived for the case $F \gg 1$.

Two-dimensional maps of the liquid surface elevation are shown in Fig. 5 for the same Froude numbers. As can be seen in Figs. 4 and 5, for the values of the Froude number $F = 0.3$ and less, the scale $h$ of the depth of the ball immersion dominates in the shape of the liquid surface (Figs. 4a and 5a). However, already at $F = 0.45$ (Fig. 4b and 5b), when the wavelength $\lambda_g = 2\pi V^2/g$, corresponding to the Cherenkov resonance, exceeds the depth of immersion $h$, the resonant ship waves begin to dominate on the surface elevation profile. At $F \sim 1$, the Kelvin wedge is clearly formed (Fig. 5c and 5d). It consists of two types of waves. Lateral waves are concentrated at the edges of the wedge. Their fronts are approximately parallel to the edges of the wedge. Transverse waves extend inside the wedge. Their fronts are perpendicular to the trajectory of the ball motion. With further increase in the Froude number the dominant wavelength gradually increases, and the surface disturbances are more and more concentrated near the edge of the wedge (Fig. 5e and 5f), the transverse waves gradually disappear, and the angle at the Kelvin wedge top is reduced.

## VI. VERTICAL OSCILLATION OF A SUBMERGED BALL

In this section, we assume that the ball performs small vertical harmonic oscillation, so that

$$X(t) = Y(t) = 0, \qquad Z(t) = -h + \delta Z\cos(\omega t).$$

We expand Eq. (35) over vibration amplitude $\delta Z$, assuming it to be small compared to the characteristic scales, which give the main contribution to the integral (34):

$$\delta Z \ll (h, g/\omega^2), \tag{70}$$

Then,

$$\phi_k^{(0)}(t) = \pi a^3\,\mathrm{e}^{-kh}\,\delta Z\,\omega\sin(\omega t). \tag{71}$$

Putting this expression into the integral (39), we again multiply it by the factor $\mathrm{e}^{\mu t}$ to simulate the assumption of the ball resting in the limit $t \to -\infty$. Integrating, we find

$$\zeta_k(t) = -\pi a^3 k\,\mathrm{e}^{-kh}\,\frac{\omega\,(\omega + i\mu)\,\mathrm{e}^{-i\omega t}}{kg - (\omega + i\mu)^2}\,\mathrm{e}^{\mu t}\,\delta Z + \text{c.c.}, \tag{72}$$

where c.c. stands for a complex conjugate term. After transition to polar coordinates in Eq. (20) using the formulas $k_y = k\cos\theta$ and $\varkappa = k\sin\theta$, integrating with respect to $\theta$ and passing to the limit $\mu \to 0+$, we obtain

$$\xi(r, t) = \operatorname{Re}[\bar{\xi}(r)\,\mathrm{e}^{-i\omega}], \tag{73}$$

where

$$\bar{\zeta}(r) = -\frac{a^3\omega^2}{g}\delta Z\int_0^\infty \frac{k^2\,\mathrm{J}_0(kr)\,\mathrm{e}^{-hk}}{k - (\omega + i0)^2/g}\,\mathrm{d}k, \tag{74}$$

is the complex amplitude, $\mathrm{J}_0$ denotes the Bessel function of zero order, and $r = \sqrt{x^2 + y^2}$.

In the limit of low frequencies, when the $\omega^2 \ll g/x$, we neglect the term $(\omega + i0)^2/g$ in the denominator of the integrand; then the integral can be calculated in a general form. Contribution of the singularity in the integrand can be taken into account by adding a half of the residue (multiplied by $2\pi i$) at the point $q = (\omega + i0)^2/g$:

$$\zeta = -\frac{a^3 h\omega^2/g}{(h^2 + r^2)^{3/2}}\,\delta Z\cos(\omega t) - \\ -\frac{\pi a^3\omega^2}{g}\delta Z\left[k^2\,\mathrm{J}_0(kr)\,\mathrm{e}^{-kh}\right]_{k=\omega^2/g}\sin(\omega t). \tag{75}$$

In this approximation, the added half-residue is exponentially small so that the small harmonic oscillation of the submerged ball creates an almost standing wave on the liquid surface directly above the ball, and the wave amplitude increases in proportion to $\omega^2$.

In the high frequency limit, when the $\omega^2 \gg g/h$, in the denominator of the integrand, on the contrary, we keep only the term $(\omega + i0)^2/g$; then, the integral is again calculated in a general form. Contribution of the singularity in the integrand can be taken into account by adding a half of the residue at $q = (\omega + i0)^2/g$ multiplied by $2\pi i$:

$$\zeta = \frac{a^3(2h^2 - r^2)}{(h^2 + r^2)^{5/2}}\,\delta Z - \\ - \frac{\pi a^3 \omega^2}{g}\delta Z \left[k^2\, \mathrm{J}_0(kr)\, \mathrm{e}^{-kh}\right]_{k=\omega^2/g} \sin(\omega t). \quad (76)$$

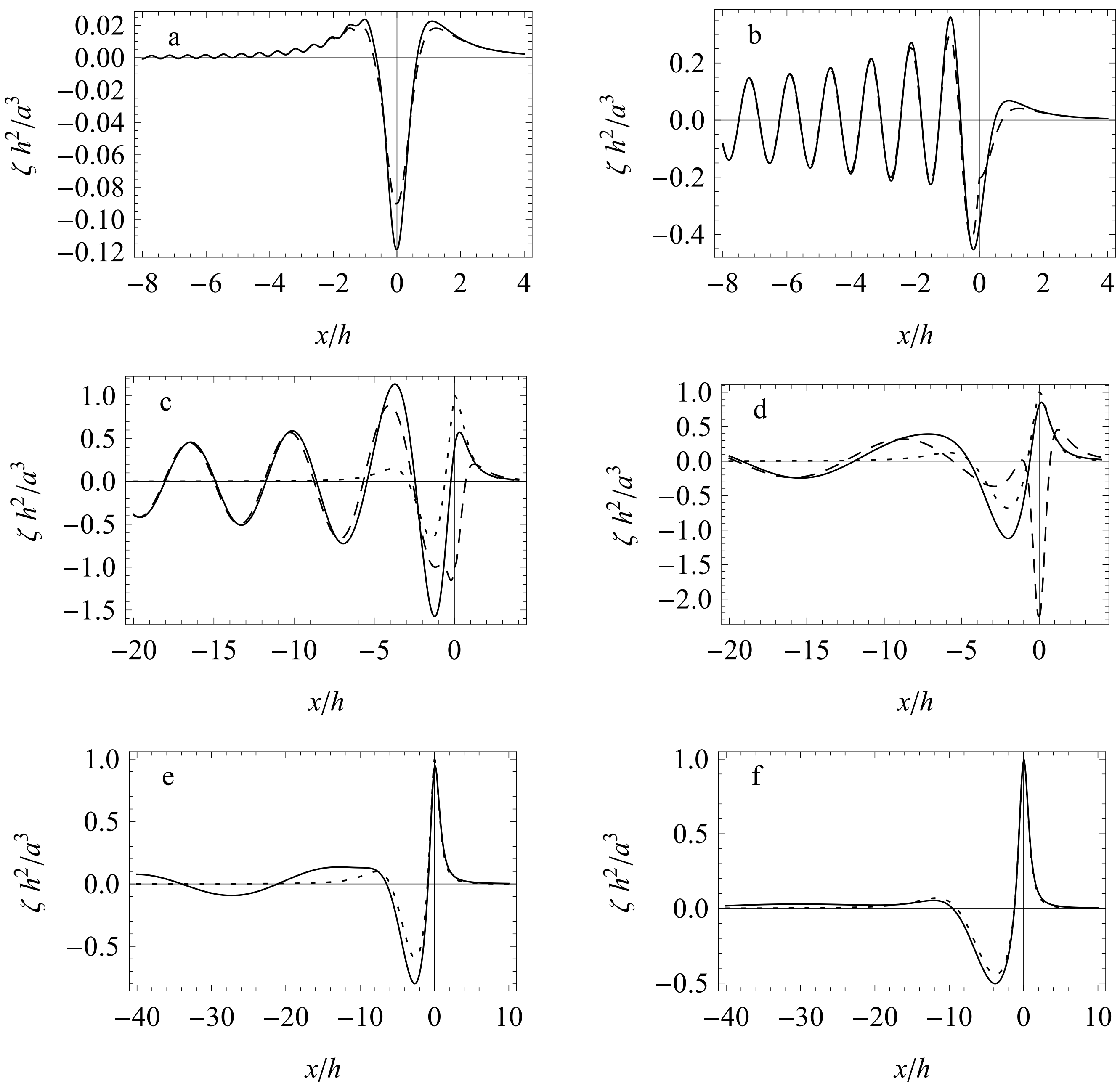

Figure 4. Profile of ship waves in the central section $y = 0$ for different values of the Froude number: a) $F = 0.3$, $\lambda_g/h = 0.57$; b) $F = 0.45$, $\lambda_g/h = 1.27$; c) $F = 1.0$, $\lambda_g/h = 6.28$; d) $F = 1.5$, $\lambda_g/h = 14.13$; e) $F = 2.0$, $\lambda_g/h = 25.1$; f) $F = 3.0$, $\lambda_g/h = 56.55$; exact formulas (52), (54) and (57) (solid line), approximate formulas (58) and (64) (dashed line) for the case $F \ll 1$, approximate formulas (66) and (68) (dotted line) for the case $F \gg 1$.

In this approximation, the addition of the half-residue is also small and oscillation of the ball again creates an almost standing wave on the liquid surface directly above the ball with the amplitude of the wave, which does not depend on $\omega$.

We also give a formula for the complex amplitude of the vibrations at $r = 0$:

$$\overline{\zeta}(0) = \frac{a^3 \delta Z \omega^6}{g^3}\left[e^{-h\omega^2/g}\left(\mathrm{Ei}\left(h\omega^2/g\right) - i\pi\right) - \\ - \frac{g^2 + gh\omega^2}{h^2\omega^4}\right]. \quad (77)$$

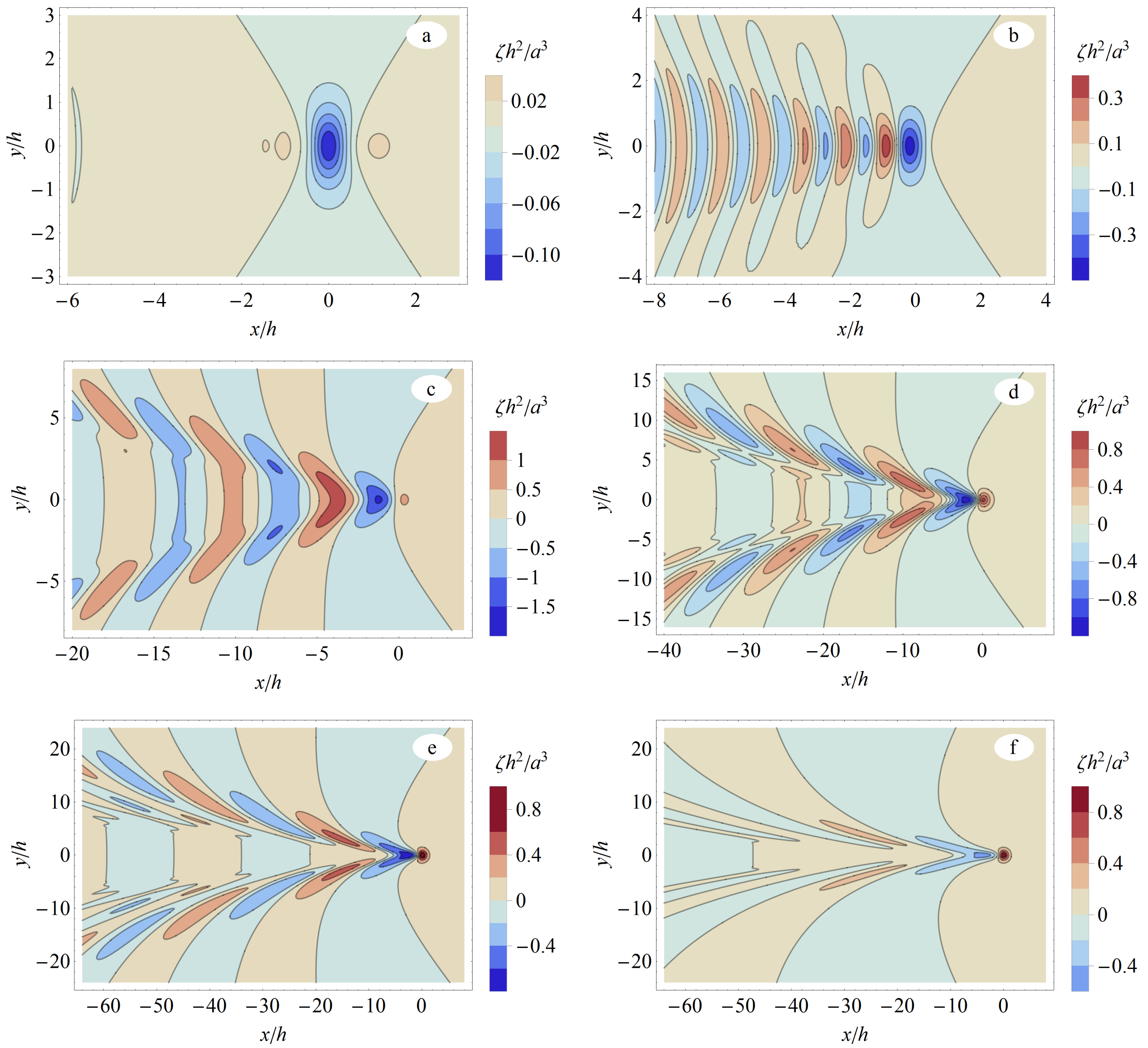

Figure 5. (Color online) Profile of the ship wave at different values of the Froude number: a) $F = 0.3$, $\lambda_g/h = 0.57$; b) $F = 0.45$, $\lambda_g/h = 1.27$; c) $F = 1.0$, $\lambda_g/h = 6.28$; d) $F = 1.5$, $\lambda_g/h = 14.13$; e) $F = 2.0$, $\lambda_g/h = 25.13$; f) $F = 3.0$, $\lambda_g/h = 56.55$.

Here

$$\mathrm{Ei}(z) = -\int_{-z}^{\infty} \frac{\mathrm{e}^{-t}}{t}\,\mathrm{d}t$$

denotes the exponential integral function, which is defined as the principal value of the above integral.

As follows from Eq. (77), the amplitude of oscillations of the liquid surface at the epicenter (ie, at a point above the center of the ball) after normalization over the parameter $(a/h)^3\delta Z$ depends only on dimensionless parameter $\omega^2 h/g$. This dependency is shown in Fig. 6. Using the expansion

$$\mathrm{Ei}(z) \approx \gamma + \ln(z) + z + \frac{z^2}{4} + \ldots$$

for $0 < z \ll 1$, we find

$$\overline{\zeta}(0) \approx -\frac{a^3\omega^2}{h^2 g}\delta Z \tag{78}$$

for $h\omega^2/g \ll 1$. Using the expansion

$$\mathrm{e}^{-z}\,\mathrm{Ei}(z) \approx \frac{1}{z} + \frac{1}{z^2} + \frac{2}{z^4} + \ldots$$

for $x \to \infty$ gives

$$\overline{\zeta}(0) \approx \frac{2a^3}{h^3}\delta Z. \tag{79}$$

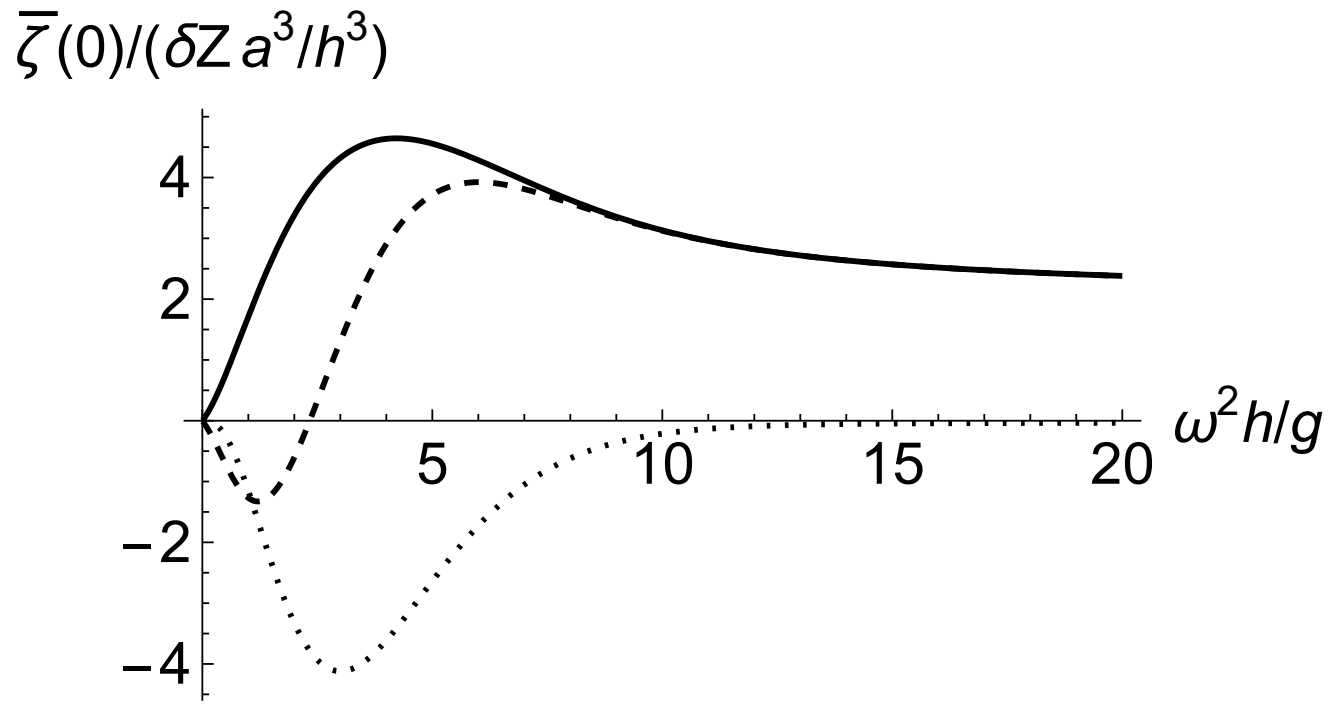

Figure 6. Absolute magnitude (solid line), real (dashed line) and imaginary (dotted line) parts of the complex amplitude $\overline{\zeta}(0)$ of oscillations of the liquid surface at the epicenter above the ball versus dimensionless parameter $\omega^2 h/g$.

The expressions (78) and (79) confirm respectively Eqs. (75) and (76).

The results of numerical calculation of the radial profile of the vertical oscillation on the liquid surface by Eq. (74) are shown in Fig. 7. From a comparison of the profiles a, b, c, d for different values of the dimensionless parameter $\omega^2 h/g$ is clearly seen that near the center of the oscillation picture an almost standing wave is excited in the limit of small and large values. Waves traveling from the epicenter are noticeable at intermediate values of the parameter $\omega^2 h/g$, close to unity, as in Fig. 7b and 7c. This is clearly seen in attached animations.

## VII. CONCLUSION

In this paper we have proposed a method for solving the non-stationary problem of excitation of ship waves by underwater object, which moves with an arbitrary velocity in a non-viscous fluid. The vertical displacement of the liquid surface, produced by the underwater ball-shaped object, is given by Eq. (39). Equation (38) can be used to construct a solution for a submerged object other than a ball. A corresponding function $\phi_k^{(0)}(\tau)$ should be computed as a solution of the problem about motion of such an object in unbounded non-viscous liquid. For example, this function can be easily found for a prolate ellipsoid. Main difficulty in this case is transferred to computation of the integrals in Eqs. (38) and (20).

For comparison with previously published results we considered an example of a ball which moves at a constant velocity along a straight path parallel to the liquid surface at a depth large in comparison with the ball radius. Vertical elevation of the liquid surface in this case is given by Eq. (54). It contains two terms $\zeta_0$ and $\zeta_1$, which are expressed by single integrals (57) and (52), respectively. These terms are derived by means of deforming the path of integration in the complex plane, which we call the Peters transformation. It removes the poles in the integrand in intermediate calculations, the treatment of which is equivalent to the Landau bypass rule known in the theory of plasma waves in the plasma. In the limit of both small and large values of the Froude number $F = V/\sqrt{gh}$ we derived asymptotic expression for $\zeta_0$ and $\zeta_1$. It has been found that in both cases the first term $\zeta_0$, which describes the Bernoulli hump, generally predominates with respect to the second term $\zeta_1$, which describes the Kelvin wedge. However, this second term $\zeta_1$ is comparable with the first one at $F \sim 1$. Moreover, $|\zeta_1| \gg |\zeta_0|$ at the edges of the Kelvin wedge if $F \gg 1$.

It has been noted that the previously known Havelock's solution also contains two terms, one of which is expressed by a rational fraction, and the other contains a two-dimensional integral. These terms have no clear physical meaning. In particular, they do not vanish when the ball speed tends to zero, whereas the elevation of the liquid surface in this case should be zero everywhere. In other words, the shape of the liquid surface is calculated as the result of the almost total reduction of two large terms.

Comparison of the results of calculation of the vertical displacement of the liquid surface for different values of the Froude number shows qualitative difference of the surface shape for large and small values of $F$. In particular, it has been found that the Kelvin wedge is not formed if $F \lesssim 0.3$. We have confirmed the conclusion that the angle at the apex of the Kelvin wedge decreases as $F$ increases provided that $F > 1$.

As an example of non-uniform motion we have solved the problem of the excitation of ship waves by vertical oscillation of the submerged ball. For this case we have obtained estimations of the amplitude of elevation of the liquid surface at high and low frequency of oscillation of the ball. Exact solution of this problem has been obtained in the linear approximation, which reduces to the calculation of one-dimensional integral.

Our proposed method of solving the problem of liquid surface waves excitation by a submerged object, which moves at a variable speed, can be used to test numerical methods applied for similar tasks.

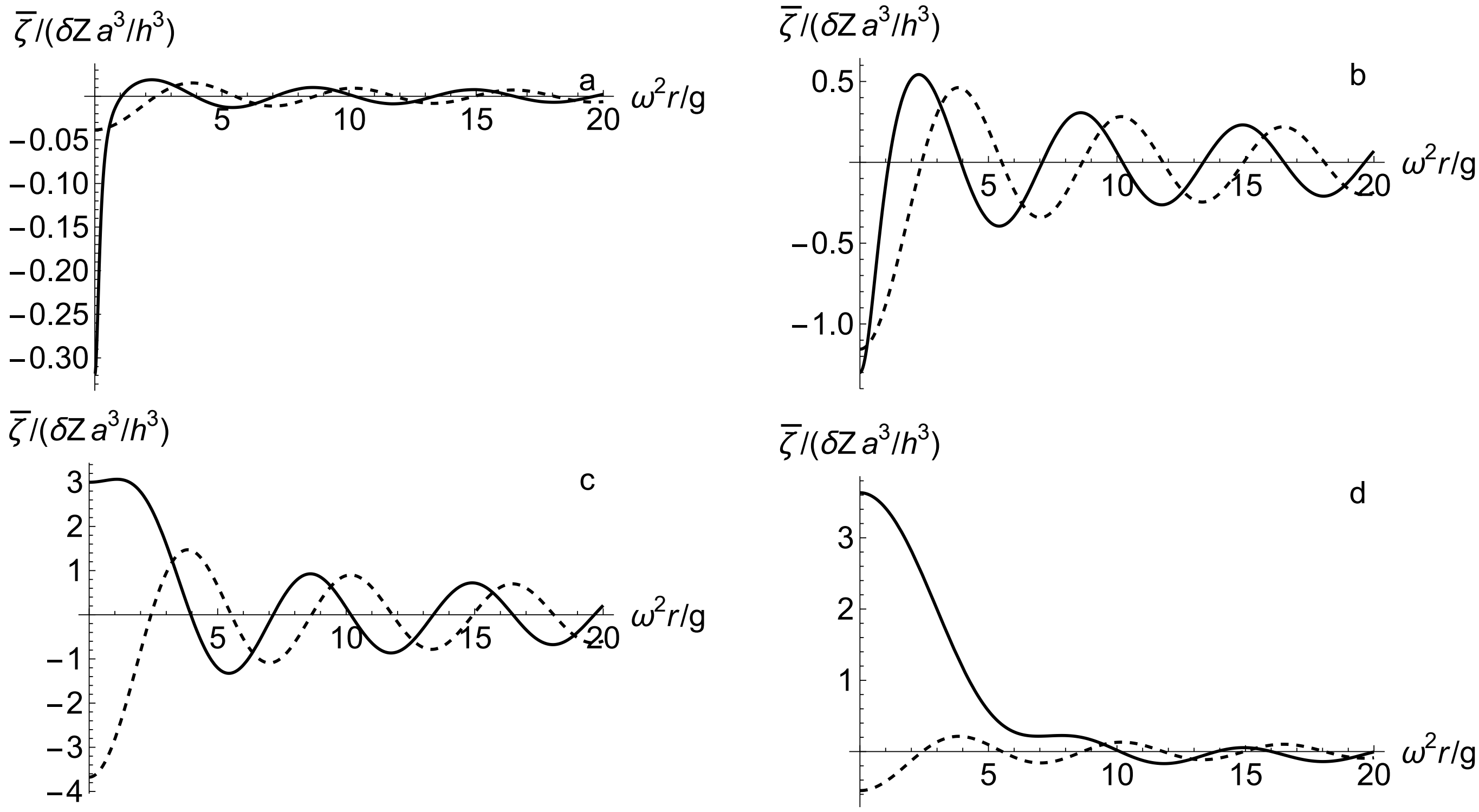

Figure 7. (Video online) Radial profiles of the liquid surface oscillation caused by vertical vibration of the submerged ball for different values of the parameter $\omega^2 h/g$: (a) $\omega^2 h/g = 0.25$; (b) $\omega^2 h/g = 1$; (c) $\omega^2 h/g = 4$; (d) $\omega^2 h/g = 8$. Click on the image to watch the video (available in Adobe Reader). Static images in the printed version of the paper show real (solid line) and imaginary (dashed line) parts of the complex amplitude $\overline{\zeta}$, Eq. (74). In video, blue line shows running wave given by exact Eqs. (73) and (74), and burgundy line shows approximate solution given by Eqs. (75) and (76).

Finally, we enlist some of the applicability conditions for the linear approximation used in this paper.

One condition should be deduced from the comparison of the term $v^2/2 \sim \zeta^2\omega^2/2$ in the boundary condition (7) compared with $g\zeta$, namely: $\zeta \ll 2g/\omega^2 \sim 2g/kg \sim \lambda$.

The second condition is $a \ll h$. It is necessary in order to justify the fact that we have neglected the contribution of the potential of the gravity wave $\phi_1$ in the boundary conditions on the surface of the ball.

### Appendix A: The Havelock solution

Let's transform Eq. (39) to Eq. (24) from Havelock's paper [8]:

$$\zeta(x,y,t) = \frac{a^3 h}{[x^2+y^2+h^2]^{3/2}} - a^3 \int_0^\infty \mathrm{d}u\, \mathrm{e}^{-\mu u/2} \times \\ \times \int_0^\infty \mathrm{d}k k\sqrt{kg}\, \mathrm{e}^{-kh} J_0\left(k\sqrt{(x+cu)^2+y^2}\right) \sin\left[\sqrt{kg}u\right], \tag{A1}$$

where $c = -V$ and $\mu \to 0+$. To accomplish this, we introduce function

$$\Phi_k^{(0)}(t) = -\frac{\pi a^3}{k}\, \mathrm{e}^{kZ(t)-ik_xX(t)-ik_yY(t)}\, \mathrm{e}^{\mu t/2}, \tag{A2}$$

with $\mu > 0$, such that

$$\frac{\partial}{\partial t}\Phi_k^{(0)}(t) = \phi_k^{(0)}(t)$$

in the limit $\mu \to 0$, see Eq. (35). Function $\zeta_k(t)$ can be expressed through $\Phi_k^{(0)}$, if one more integration by parts is done in Eq. (39):

$$\zeta_k(t) = -2k\Phi_k^{(0)}(t) - \\ - 2k\sqrt{kg}\int_{-\infty}^t \sin\left[\sqrt{kg}(\tau - t)\right]\Phi_k^{(0)}(\tau)\,\mathrm{d}\tau. \tag{A3}$$

Factor $\mathrm{e}^{\mu t/2}$ in (A2) ensures that function $\Phi_k^{(0)}(t)$ tends to zero as $t \to -\infty$, which justifies the integration by parts.

Since Havelock considered the case of rectilinear motion of the ball at a given constant depth, we assume for brevity that $Y(z) = 0$, $Z(y) = -h$ and calculate $\zeta(x,y,t)$ by performing inverse Fourier transform (20). In this case the first term in (A3) yields the expression

$$\iint \frac{\mathrm{d}k_x}{2\pi}\frac{\mathrm{d}k_x}{2\pi}\,\mathrm{e}^{ik_x x+ik_y y}\left(-2k\Phi_k^{(0)}(t)\right) = = \frac{a^3 h\,\mathrm{e}^{\mu t/2}}{[(x-X(t))^2+y^2+h^2]^{3/2}},$$

which coincides with the first term in Eq. (A1) for $\mu = 0$, if we pass to the reference frame of moving ball (in this frame $X(t) = 0$). Performing Fourier transform of the second term, we change the integration variables $k_x = k\cos(\theta)$, $k_y = k\sin(\theta)$. Then, the integral over $\theta$ is expressed through the Bessel function

$$J_0(\alpha) = \frac{1}{2\pi}\int_{-\pi}^{\pi}\mathrm{e}^{i\alpha\cos(\theta)}\,\mathrm{d}\theta.$$

As a result, the second summand in Eq. (A3) gives the term

$$\iint \frac{\mathrm{d}k_x}{2\pi}\frac{\mathrm{d}k_x}{2\pi}\,\mathrm{e}^{ik_x x+ik_y y}\left(-2k\sqrt{kg}\int_{-\infty}^{t}\sin\left[\sqrt{kg}(\tau-t)\right]\Phi_k^{(0)}(\tau)\,\mathrm{d}\tau\right) = = a^3\int_{-\infty}^{t}\mathrm{d}\tau\,\mathrm{e}^{\mu\tau}\int_0^{\infty}\mathrm{d}k k\sqrt{kg}\,\mathrm{e}^{-kh}J_0\left(k\sqrt{(x-X(\tau))^2+y^2}\right)\sin\left[\sqrt{kg}(\tau-t)\right],$$

which coincides with the second term in Eq. (A1), if we put $X(\tau) = V\tau$, $t = 0$ and make the substitution $\tau \to -u$.